\documentclass[11pt]{article}
\usepackage{amssymb}
\usepackage{graphicx}
\usepackage{layout}
\usepackage{bm}
\begin{document}

\title{Gravito-electromagnetic Aharonov-Bohm effect: some rotation effects  revised}
\author{Matteo Luca Ruggiero$^{\S,\P}$
\\
\small
$^\S$ Dipartimento di Fisica, Politecnico di Torino,\\
\small $^\P$ INFN, Sezione di Torino\\
\small e-mail: matteo.ruggiero@polito.it} \maketitle

\begin{abstract}
By means of the description of the standard relative dynamics in
terms of gravito-electromagnetic fields, in the context of natural
splitting,  we formally introduce  the gravito-magnetic
Aharonov-Bohm effect. Then, we interpret the Sagnac effect as a
gravito-magnetic Aharonov-Bohm effect and we exploit this
formalism for studying the General Relativistic corrections to the
Sagnac effect in stationary and axially symmetric geometries.
\end{abstract}

\section{Introduction}\label{sec:intro}

In a previous paper\cite{rizzi03a} we developed a formal approach,
for the description of the relativistic dynamics, which lead to
what we called \textit{gravito-magnetic Aharonov-Bohm effect}.
This approach is based on the description of the standard relative
dynamics in the context of the \textit{natural splitting},
introduced by Cattaneo and subsequently developed by himself,
Ferrarese and collaborators.\footnote{See \cite{rizzi_ruggiero04},
and references therein.} The formulation that we gave of the
gravito-magnetic Aharonov-Bohm effect holds in exact theory, and
it generalizes some old results obtained at first order
approximation\cite{sakurai80}. In the previous paper we used this
approach to calculate the Sagnac time delay, for matter or light
beams counter-propagating in a rotating reference frame in flat
space-time. In doing so, we exploited the formal analogy between
matter (or light) beams counter propagating in circular
trajectories in gravitational or inertial fields, and charged
beams propagating in a region where a magnetic potential is
present. Here we apply the formalism of the gravito-magnetic
Aharononv-Bohm effect for studying the Sagnac effect in curved
space-time. Tartaglia\cite{tartaglia98} studied the General
Relativistic corrections to the Sagnac effect in Kerr space-time,
while an approach similar to ours, i.e. based on a
gravito-magnetic Aharonov-Bohm effect, was developed by Cohen and
Mashhoon\cite{cohen93}, who studied the gravito-magnetic time
delay in the weak gravitational field of a rotating source, at
first order approximation. Our formalism allows to generalize
these results to arbitrary stationary and axially symmetric
geometries, in full theory without any approximations.

The paper is organized as follows: in Section \ref{sec:gem_desc}
we recall the main features of the gravito-magnetic description of
dynamics, in Sections \ref{sec:ab} and \ref{sec:gem_ab} we
illustrate, respectively, the Aharonov-Bohm effect and the
gravito-magnetic Aharonov-Bohm effect. In Section \ref{sec:sagnac}
we study the Sagnac effect in flat and curved space-time.

\section{Gravito-electromagnetic description of relativistic
dynamics}\label{sec:gem_desc}

By applying Cattaneo's splitting, the dynamics of  massive or
massless particles, relative to a given time-like congruence
$\Gamma$ of unit vectors $\bm{\gamma}(x)$, can be described in
terms of a "gravito-electromagnetic" analogy. Here we briefly
review the main features of this approach, in order to formulate,
in the following Sections, the gravito-magnetic Aharonov-Bohm
effect. The details of the calculations may be found in
\cite{rizzi_ruggiero04}.

The congruence $\Gamma$ defines the physical reference frame with
respect to which the particles move, and an observer at rest on
this physical frame has a world-line which coincides with one of
the world-lines of the congruence.

Let $\gamma^\mu$ be the components of the unit vectors field
$\bm{\gamma}(x)$ in coordinates adapted to the
congruence.\footnote{Greek indices run from 0 to 3, Latin indices
run from 1 to 3, the signature of the space-time metric is
$(-1,1,1,1)$.} Their expression in terms of the metric tensor
$g_{\mu\nu}$ is
\begin{equation}
\left\{
\begin{array}{l}
\gamma ^{{0}}=\frac{1}{\sqrt{-g_{{00}}}} \\
\gamma ^{i}=0
\end{array}
\right. \;\;\;\;\;\;\;\;\;\;\;\;\left\{
\begin{array}{l}
\gamma _{0}=-\sqrt{-g_{{00}}} \\
\gamma _{i}=g_{i{0}}\gamma ^{{0}}
\end{array}
\right.  \label{eq:gammas11}
\end{equation}

Then, if we introduce the \textit{gravito-electric potential}
$\phi^G$ and the \textit{gravito-magnetic potential}
$\tilde{A}_i^G$, defined by

\begin{equation}
\left\{
\begin{array}{c}
\phi ^{G}\doteq -c^{2}\gamma ^{0} \\
\widetilde{A}_{i}^{G}\doteq c^{2}\frac{\gamma _{i}}{\gamma _{0}}
\end{array}
\right.   \label{eq:gengravpot11}
\end{equation}

we can write the \textit{gravito-magnetic field}

\begin{equation}
\widetilde{B}_{G}^{i}\doteq \left( \widetilde{{\bf \nabla }}\times
\bm{\tilde{A}}_{G}\right) ^{i}  \label{eq:gengravmag}
\end{equation}

and the \textit{gravito-electric field}:
\begin{equation}
\widetilde{E}^{G}_i\doteq -\left( -\widetilde{\partial} _{i}\phi
_{G}-\partial _{0}\widetilde{A}^{G}_i\right)  \label{eq:egen}
\end{equation}

in analogy with the definitions of classical electromagnetism.
Notice that we used the following differential operator
\begin{equation}
\tilde{\partial}_\mu \doteq \partial_\mu+ \gamma_\mu \gamma^0
\partial_0 \label{eq:tranpartdev}
\end{equation}
which is called \textit{transverse partial derivative}, and it
defines the space-projection of the local gradient; furthermore,
we recall here that the "\ $\tilde{}$\ " denotes
\textit{space-vectors}, i.e. vectors that are obtained by
projecting the world-vectors onto the 3-dimensional subspace
orthogonal to the time-like direction spanned by $\gamma^\mu$ (see
the following
Remark).\\

\small \textbf{Remark \ }Let $\mathcal{M}^{4}$ be  a (pseudo)riemannian manifold $%
\mathcal{M}^{4}$, that is a pair $\left(
\mathcal{M},\mathbf{g}\right)$, where $\mathcal{M}$ is a connected 4-dimensional Haussdorf manifold and $%
\mathbf{g}$ is the metric tensor.\footnote{%
The riemannian structure implies that $\mathcal{M}$ is endowed
with an affine connection compatible with the metric, i.e. the
standard Levi-Civita connection.} $\mathcal{M}^{4}$ is the
mathematical model of the physical space-time. At each point $p\in
\mathcal{M}^4$, the tangent space $T_{p}$ can be split into the
direct sum of two subspaces: $\Theta _{p}$, spanned by $\gamma
^{\mu }$, which we shall call \textit{local time direction} of the
given frame, and $\Sigma _{p}$, the 3-dimensional subspace which
is supplementary (orthogonal) with respect to $\Theta_{p}$;
$\Sigma _{p}$ is called \textit{local space platform} of the given
frame. So, the tangent space can be written as the direct sum
\begin{equation}
T_{p}=\Theta _{p}\oplus \Sigma _{p}  \label{eq:tangsum}
\end{equation}
Let $\{\bm{e}_\mu\}$ be a basis of $T_p$. A vector $\bm{v}=v^\mu \bm{e}_\mu    \in T_{p}$ can be projected onto $\Theta _{p}$ and $%
\Sigma _{p}$ using the \textit{time projector}
\begin{equation}
P_\Theta \doteq -\gamma _{\mu }\gamma _{\nu } \label{eq:ptheta1}
\end{equation}
and the \textit{space projector}
\begin{equation}
P_\Sigma \doteq \gamma _{\mu \nu }\doteq g_{\mu \nu }+\gamma _{\mu
}\gamma _{\nu } \label{eq:psigma1}
\end{equation}
in the following way:
\begin{equation}
\left\{
\begin{array}{rclll}
\bar{v}_{\mu } & \equiv & P_{\Theta }\left( \,v_\mu\right) &
\doteq &
-\gamma _{\mu }\gamma _{\nu }v^{\nu } \\
\widetilde{v}_{\mu } & \equiv & P_{\Sigma }\left( \,v_\mu\right) &
\doteq & \gamma _{\mu \nu }v^{\nu }=\left( g_{\mu \nu }+\gamma
_{\mu }\gamma _{\nu }\right) v^{\nu }= v_{\mu }+v^{\nu }\gamma
_{\nu }\gamma _{\mu }
\end{array}
\right.  \label{eq:DefSplitComp}
\end{equation}
From Eq. (\ref{eq:DefSplitComp}), $\forall \bm{v} \in T_p$ we have
\begin{equation}
v_\mu=\bar{v}_\mu+\widetilde{v}_\mu=
P_\Theta(v_\mu)+P_\Sigma(v_\mu) \label{eq:projectspacetime3}
\end{equation}
This defines the \textit{natural splitting} of a vector
$\bm{v}$. $\square$\\

\normalsize

In terms of the physical quantities that we have introduced so
far, the space projection of the geodesics equation for matter or
massless particles can be written in the form\footnote{Notice that
here we consider the covariant expression of the space projection
of the equation of motion; in general, the contravariant
expression has en extra-term which contains the Born tensor (see
\cite{ferrarese}, Sec. 6.5). However, the latter is null when the
space-time metric does not depend on the time coordinate, which is
the case of the metrics we are interested in. Hence, in what
follows, we may consider the covariant expression without loss of
generality.}

\begin{equation}
\frac{\hat{D}\widetilde{p}_{i}}{dT}=m \widetilde{E}^{G}_i+m\gamma
_{0}\left( \frac{\bm{\tilde{v}}}{c}\times
\bm{\tilde{B}}_{G}\right) _{i} \label{eq:motogen1}
\end{equation}

For  particles of rest mass $m_0$, $m$ is defined by
\begin{equation}
m\doteq\frac{m_0}{\sqrt{1-\frac{\widetilde{v}^2}{c^2}}}
\label{eq:relmass}
\end{equation}
and it is the \textit{standard relative mass}, which depends on
their \textit{standard relative velocity} $\bm{\tilde{v}}$. For
massless particles (such as photons) $m$ is proportional to the
\textit{standard relative energy}, and it is defined by
\begin{equation}
m=\frac{h\nu}{c^2} \label{eq:mhnu}
\end{equation}
in terms of the frequency $\nu$ and the Planck constant. In Eq.
(\ref{eq:motogen1}) $dT$ is the \textit{standard relative time
interval}
\begin{equation}
dT=-\frac{1}{c}\gamma_\mu dx^\mu \label{eq:dT1}
\end{equation}
and it represents the proper time interval as measured by an
observer at rest in $\Gamma$.\\

Eq. (\ref{eq:motogen1}) shows that the variation of the
\textit{standard relative momentum} vector $\tilde{p}_i$ (in terms
of a suitable derivative operator $\frac{\hat{D}}{dT}$) is
determined by the action of a \textit{gravito-electromagnetic
Lorenz force}. By means of Eq. (\ref{eq:motogen1})  we may give
 a description of the relativistic dynamics in
analogy with the electromagnetic theory. This description has been
obtained without any approximations, consequently it applies to
arbitrary congruences, in both flat and curved space-time. Of
course, if one performs a first order approximation, the
gravito-electromagnetic analogy that we have obtained corresponds
to the well known gravito-electromagnetic description of the weak
field approximation of General Relativity (see for instance
\cite{mashhoon01},\cite{ruggiero02},\cite{mashhoon03}).

If the particles are not free, but are acted upon by an external
force field $F^\alpha$, Eq. (\ref{eq:motogen1}) becomes

\begin{equation}
\frac{\hat{D}\widetilde{p}_{i}}{dT}=m \widetilde{E}^{G}_i+m\gamma
_{0}\left( \frac{\bm{\tilde{v}}}{c}\times
\bm{\tilde{B}}_{G}\right) _{i} + \widetilde{F}_i
\label{eq:motogenF}
\end{equation}
where the space projection $\widetilde{F}_i$ of the external field
has been introduced. For instance, the constraints that force a
particle to move along a circular path (or guide) are external
fields: this is what happens to beams of particles that propagate
in a rotating ring interferometer, which is the typical situation
of the Sagnac effect.

We recall here that the gravito-magnetic effects are related to
the rotational features of the congruence we deal with. In fact,
starting from the definition of \textit{space vortex tensor} of
the congruence $\Gamma$
\begin{equation}
\widetilde{\Omega }_{\mu \nu } =\gamma _{0}\left[ \widetilde{\partial }%
_{\mu }\left( \frac{\gamma _{\nu }}{\gamma _{0}}\right)
-\widetilde{\partial }_{\nu }\left( \frac{\gamma _{\mu }}{\gamma
_{0}}\right) \right] \label{eq:vortex}
\end{equation}
we may define and axial 3-vector $\tilde{\omega}^i$, by means of
the relation
\begin{equation}
\widetilde{\omega} ^{i}\doteq \frac{c}{4}\varepsilon ^{ijk}\tilde{\Omega}_{jk}=\frac{c}{2%
}\varepsilon ^{ijk}\gamma _{0}\tilde{\partial}_{j}\left( \frac{\gamma _{k}}{%
\gamma _{0}}\right)   \label{eq:vettorevortice}
\end{equation}
where
\begin{displaymath}
\varepsilon^{ijk} \doteq \frac{1}{\sqrt{det(\gamma
_{ij})}}\epsilon ^{ijk}
\end{displaymath}
is the \textit{Ricci-Levi-Civita tensor density}, defined in terms
of the completely antisymmetric \textit{Ricci-Levi-Civita tensor}
$\epsilon^{ijk}$ and of the space metric tensor $\gamma_{ij}$. The
gravito-magnetic field $\widetilde{B}_{G}^{i}$ is proportional to
this vector:
\begin{equation}
\widetilde{\omega}^{i}= \frac{1}{2c}\gamma
_{0}\widetilde{B}_{G}^{i} \label{eq:omega2}
\end{equation}

This means that, for instance, when we deal with a congruence of
rotating observers, or when we deal with a congruence of observers
around a rotating source of gravitational field, we do expect
gravito-magnetic effects to appear.

\section{The Aharonov-Bohm effect}\label{sec:ab}

\begin{figure}[top]
\begin{center}
\includegraphics[width=7cm,height=7cm]{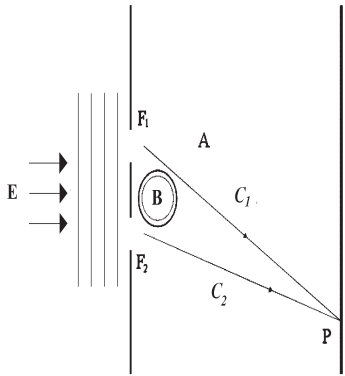}
\caption{\small A single coherent charged beam, originating in
$E$, is split into two parts (passing through the two slits $F_1$
and $F_2$) that propagate, respectively,  along the paths $C_1$
and $C_2$ (in the figure these paths are represented,
respectively, by  $EF_1P$ and $EF_2P$). Between the paths a
solenoid is present; the magnetic field $\bm{\vec{B}}$ is entirely
contained inside the solenoid, while outside there is a constant
vector potential $\bm{\vec{A}}$. In $P$, the beams interfere and
an additional phase shift, provoked by the magnetic field confined
inside the solenoid, is observed.} \label{fig:ab2}
\end{center}
\end{figure}
\normalsize

Consider the two slits experiment (see Figure \ref{fig:ab2}) and
imagine that a single coherent charged beam is split into two
parts, which travel in a region where only a magnetic field is
present, described by the 3-vector potential
$\bm{\vec{A}}$;\footnote{Boldface arrowed letters refers to
3-vectors in flat space.} then the beams are recombined to observe
the interference pattern. The phase  of the two wave functions, at
each point of the pattern, is modified, with respect to the case
of free propagation ($\bm{\vec{A}}~=~0$), by the magnetic
potential. The magnetic potential-induced phase shift has the
form\cite{aharonovbohm59}
\begin{equation}
\Delta \Phi=\frac{e}{c\hbar }\oint_{C}\bm{\vec{A}}\cdot {\rm d}\bm{\vec{x}}=\frac{e}{%
c\hbar }\int_{S}\bm{\vec{B}}\cdot {\rm d}\bm{\vec{S}}
\label{eq:ab}
\end{equation}
where $C$ is the oriented closed curve, obtained as the sum of the
oriented paths $C_{1}$ and $C_{2}$ relative to each component of
the beam. Eq. (\ref{eq:ab}) expresses (by means of  Stoke's
Theorem) the phase difference in terms of the flux of the magnetic
field across the surface $S$ enclosed by the curve $C$. Aharonov
and Bohm\cite{aharonovbohm59} applied this result to the situation
in which the two  split beams pass one on each side of a solenoid
inserted between the paths. Thus, even if the magnetic field
$\bm{\vec{B}}$ is totally contained within the solenoid and the
beams pass through a $\bm{\vec{B}}~=~0$ region, a resulting phase
shift appears, since a non null magnetic flux is associated to
every closed path which encloses the solenoid.

Tourrenc\cite{tourrenc77} showed that no explicit wave equation is
demanded to describe the Aharonov-Bohm effect, since its
interpretation is a pure geometric one: in fact Eq. (\ref{eq:ab})
is independent of the  physical nature of the interfering charged
beams, which can be spinorial, vectorial or tensorial. So, if we
deal with relativistic charged beams, their propagation is
described by a relativistic wave equation, such as the Dirac
equation or the Klein-Gordon equation, depending on the nature of
the beams themselves. From a physical viewpoint,  spin has no
influence on the Aharonov-Bohm effect because there is no coupling
with the magnetic field which is confined inside the solenoid.
Moreover, if the magnetic field is null, the Dirac equation is
equivalent to the Klein-Gordon equation, and this is the case of a
constant potential. As far as we are concerned,  since in what
follows we neglect spin,   we shall just use Eq. (\ref{eq:ab}) and
we shall not explicitly refer  to any relativistic wave equation.

Indeed, things are different when a particle with spin, moving in
a rotating frame or around a rotating mass, is considered. In this
case a coupling between the spin and the angular velocity of the
frame or the angular momentum of the rotating mass appears (this
effect is evaluated by Hehl-Ni\cite{hehl90},
Mashhoon\cite{mashhoon88} and Papini\cite{papinilibro}).  Hence,
the formal analogy leading to the formulation of the
gravito-magnetic Aharonov-Bohm effect, which is outlined in the
following Section, holds only when the spin-rotation coupling is
neglected.

\section{The Gravito-magnetic Aharonov-Bohm effect}\label{sec:gem_ab}

In this Section, on the basis of the gravito-electromagnetic
description of the relativistic dynamics that we recalled before,
we  introduce the \textit{gravito-magnetic Aharonov-Bohm effect}.
This  enables us to outline an analogy between matter (or light)
beams, counter-propagating around circular orbits, in stationary
and axially symmetric geometries, and charged beams, propagating
in a
region where a magnetic potential is present.\\

In Eq. (\ref{eq:motogenF}) the general form of the equation of
motion, relative to a congruence $\Gamma$, is given in terms of
the gravito-electric field $\widetilde{\bm{E}}_G$, the
gravito-magnetic field $\widetilde{\bm{B}}_G$ and the external
fields. In particular, in Eq. (\ref{eq:motogenF}) a
gravito-magnetic Lorentz force appears
\begin{equation}
\widetilde{{\cal F}}''_{i}=m\gamma _{0}\left(
\frac{\bm{\widetilde{v}}}{c}\times \bm{\widetilde{B}} _{G}\right)
_{i} \label{eq:genlorentz11}
\end{equation}

By following the analogy between the magnetic and gravito-magnetic
field, we want to study the phase shift induced by the
gravito-magnetic field on  beams of matter or light particles
which, after being coherently split, make a complete round trip,
in opposite directions, along  a circumference.

The analogue of the phase shift (\ref{eq:ab}) for the
gravito-magnetic field turns out to be
\begin{equation}
\Delta \Phi =\frac{2m\gamma _{0}}{c\hbar }\oint_{C}\widetilde{\bm{A}}_{G}\cdot {\rm d}%
\widetilde{\bm{x}}=\frac{2m\gamma _{0}}{c\hbar
}\int_{S}\bm{\widetilde{B}}_{G}\cdot {\rm d}\bm{\widetilde{S}}
\label{eq:gab}
\end{equation}
where $m$ refers to the standard relative mass or standard
relative energy of the particles of the beams (see Section
\ref{sec:gem_desc}).

The phase shift (\ref{eq:gab}) has been  obtained on the basis of
the formal analogy between Eq. (\ref{eq:genlorentz11}) and  the
magnetic force:
\begin{equation}
\mathbf{F}_{Lor}=\frac{e}{c}\mathbf{v}\times \mathbf{B}
\label{eq:lorentz1}
\end{equation}
by means of the substitution
\begin{equation}
\frac{e}{c}\bm{\vec{B}} \rightarrow \frac{m \gamma_0}{c}
\widetilde{\bm{B}}_G    \label{eq:sub2}
\end{equation}
We recall here that in order to have a gravito-magnetic field, the
geometry of the congruence $\Gamma$ has to be
stationary:\footnote{We refer to the definition of stationarity
given in \cite{rizzi_ruggiero04}, since there is not common
agreement in the literature.} hence  a gravito-magnetic
Aharonov-Bohm effect arises in stationary geometries.

As we said before, the standard relative time is the proper time
for an observer or a measuring device at rest in the congruence
$\Gamma$, which constitutes the reference frame with respect to
which the beams propagate. Consequently, the proper time
difference corresponding to (\ref{eq:gab}) is obtained according
to

\begin{equation}
\Delta T = \frac{\Delta \Phi}{\omega}= \frac{\hbar}{E} \Delta \Phi
= \frac{\hbar}{mc^2}\Delta \Phi \label{eq:dtaudphi}
\end{equation}
and it turns out to be
\begin{equation}
\Delta T = \frac{2\gamma _{0}}{c^3}\oint_{C}\widetilde{\bm{A}}_{G}\cdot {\rm d}%
\widetilde{\bm{x}}=\frac{2\gamma
_{0}}{c^3}\int_{S}\bm{\widetilde{B}}_{G}\cdot {\rm
d}\bm{\widetilde{S}} \label{eq:deltatau1}
\end{equation}

We want to point out that both (\ref{eq:gab}) and
(\ref{eq:deltatau1}) have been obtained taking into account the
hypothesis that the two beams (particles) have the same velocity
(in absolute value) relative to the reference frame defined by the
congruence $\Gamma$.\footnote{Generally speaking, we deal with
wave packets, hence the velocities we refer to, here and
henceforth, are the group velocities of these wave packets (see
\cite{rizzi_ruggiero04}, Sec. 3).} In particular, if we refer to a
rotating frame in flat space-time, this hypothesis coincides with
the condition of "equal velocities in opposite directions" that we
imposed in order to obtain the Sagnac effect by using a direct
approach\cite{rizzi03b}.

As we shall see below, the time delay   (\ref{eq:deltatau1})
corresponds to the Sagnac time-delay: hence, Eq.
(\ref{eq:deltatau1}) clearly evidences the "universal" character
of the time delay Sagnac effect, since the mass (or more
correctly, the energy) of the particles of the interfering beams
does not appear. So, the Sagnac effect turns out to be an effect
of the geometry of space-time, and it can be considered universal,
in the sense that it is the same, independently of the physical
nature of the interfering beams.

\section{Sagnac effect in flat and curved space-time}\label{sec:sagnac}
\begin{figure}[top]
\begin{center}
\includegraphics[width=6cm,height=5cm]{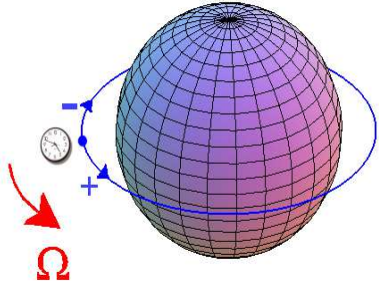}
\caption{\small We consider the  proper time difference (as
measured by a  clock rotating with constant angular velocity
$\Omega$) between the emission and absorption of the
co-propagating (+) and counter-propagating (-) beam.}
\label{fig:sphere}
\end{center}
\end{figure}
\normalsize

The Sagnac effect has been thoroughly studied in the past, and it
has been detected in many experiments (see \cite{rizzi_ruggiero04}
for a recent review). It is well known that when observing the
interference between light or matter beams (such as light beams,
electron or neutron beams and so on) counter-propagating in flat
space-time along a closed path in a rotating interferometer a
fringe shift\footnote{With respect to the interference pattern
when the device is at rest.} $\Delta \Phi$ arises. This phase
shift can be interpreted as a  time difference between the
propagation times (as measured by a clock at rest on the rotating
interferometer) of the co-rotating and counter-rotating beam. In
\cite{rizzi03a} we already showed  that the Sagnac effect, for
both matter and light beams, counter-propagating in a
interferometer rotating in flat space-time, may be obtained by
following a formal analogy with the Aharonov-Bohm effect: in a
sense, it might be thought of as a gravito-magnetic Aharonov-Bohm
effect. However, this procedure can be generalized to study the
Sagnac effect in curved space-time, in order to obtain the General
Relativistic corrections. In other words, we can study the
interference process of matter or light beams in a rotating frame
in curved space-time in terms of gravito-magnetic Aharonov-Bohm
effect. This corresponds to calculating the phase shift
(\ref{eq:gab}) and the time difference (\ref{eq:deltatau1}) as
measured, respectively, by a uniformly rotating interferometer and
by an  observer, provided with a standard clock. The time
difference corresponds to the delay between the propagation time
of the co-propagating and counter-propagating beam (see Figure
\ref{fig:sphere}). In what follows, the two beams are supposed to
have the same velocity (in absolute value) with respect to the
rotating frame.

For studying the Sagnac effect in stationary and axially symmetric
geometries,  it is sufficient to express the space-time metric in
coordinates adapted to a congruence of rotating observers.
Generally speaking the metrics we deal with are given in
coordinates adapted to a congruence of asymptotically inertial
observers; if the coordinates are spherical, the passage to a
congruence of observers uniformly rotating in the equatorial plane
is obtained by applying the (azimuthal) coordinate transformation:
\begin{equation} \left\{
\begin{array}{rcl}
ct & = & ct' \\
r & = & r' \\
\varphi & = & \varphi' -\Omega t' \\
\vartheta & = & \vartheta'
\end{array}
\right. \mathrm{\ .}  \label{eq:sagnac_trasf_cord}
\end{equation}
where $\Omega$ is the (constant) angular velocity.\\

Then, it is simple to apply the formalism that we have described
so far. The following steps give a prescription for calculating
the Sagnac effect in arbitrary stationary and axially symmetric
geometries in both flat and curved space-times.
\begin{itemize}
\item Define the time-like congruence $\Gamma$ of rotating observsers\\

\item Express the space-time metric in coordinates adapted to
congruence of rotating observers

\item Calculate the unit vectors field $\bm{\gamma}(x)$\\
\begin{displaymath}
\left\{
\begin{array}{l}
\gamma ^{{0}}=\frac{1}{\sqrt{-g_{{00}}}} \\
\gamma ^{i}=0
\end{array}
\right. \;\;\;\;\;\;\;\;\;\;\;\;\left\{
\begin{array}{l}
\gamma _{0}=-\sqrt{-g_{{00}}} \\
\gamma _{i}=g_{i{0}}\gamma ^{{0}}
\end{array}
\right.
\end{displaymath}

\item Calculate the gravito-electromagnetic potentials

\begin{displaymath}
\left\{
\begin{array}{c}
\phi ^{G}\doteq -c^{2}\gamma ^{0} \\
\widetilde{A}_{i}^{G}\doteq c^{2}\frac{\gamma _{i}}{\gamma _{0}}
\end{array}
\right.
\end{displaymath}
\item Calculate the Sagnac time delay as a gravito-magnetic
Aharonov-Bohm effect
\begin{displaymath}
\Delta T = \frac{2\gamma _{0}}{c^3}\oint_{C}\widetilde{\bm{A}}_{G}\cdot {\rm d}%
\widetilde{\bm{x}}=\frac{2\gamma
_{0}}{c^3}\int_{S}\bm{\widetilde{B}}_{G}\cdot {\rm
d}\bm{\widetilde{S}} \label{eq:deltatau11}
\end{displaymath}
\\

In the following, after reviewing the flat space-time case,   we
calculate the Sagnac effect in some curved space-times of physical
interest. We consider beams counter propagating around the source
of gravitational field,  in its  equatorial plane, along circular
orbits of radius $R=constant$.

\end{itemize}

\subsection{Minkowski space-time}\label{ssec:flat}

The flat space-time metric, in cylindrical coordinates has the
form:
\begin{equation}
ds^2=-c^2dt'^2+dr'^2+r'^2d\varphi'^2+dz'^2 \label{eq:flatmet1}
\end{equation}
By performing an azimuthal transformation
\begin{equation} \left\{
\begin{array}{rcl}
 ct& = &ct' \\
r& = &r' \\
 \varphi & = &\varphi' -\Omega t' \\
z& = &z'
\end{array}
\right. \mathrm{\ .}  \label{eq:sagnac_trasf_cord_flat}
\end{equation}
we get
\begin{equation}
ds^2=-\left(1-\frac{\Omega^2r^2}{c^2} \right)c^2dt^2+2\frac{\Omega
r^2}{c}d\varphi cdt+dr^2+r^2d\varphi^2+dz^2 \label{eq:flatmet11}
\end{equation}

Then, the non null components of the vector field \mbox{\boldmath
$\gamma$}$(x)$, evaluated along the trajectories $r=R=constant$
are
\begin{equation}
\left\{
\begin{array}{l}
\gamma ^{0} = \gamma  \\
\gamma _{0} = -\gamma ^{-1} \\
\gamma_{\varphi} =\frac{\gamma \Omega R^{2}}{c}
\end{array}
\right. \label{eq:gammasflat}
\end{equation}
where $\gamma =\left( 1-\frac{\Omega ^{2}R^{2}}{c^{2}}\right) ^{-\frac{1}{2}%
} $.
As to the gravito-magnetic potential, we have
\begin{equation}
\widetilde{A}_{2 }^{G}=\widetilde{A}_{\varphi }^{G} \doteq
c^{2}\frac{\gamma _{\varphi}}{\gamma _{0}} =-\gamma ^{2}\Omega
R^{2}c \label{eq:vectpot}
\end{equation}

Consequently, the phase shift (\ref{eq:gab}) becomes
\begin{equation}
\Delta \Phi =\frac{2m\gamma_0}{c\hbar }\int_{0}^{2\pi
}\widetilde{A}_{\varphi }^{G}d\varphi =\frac{2m}{c\hbar \gamma
}\int_{0}^{2\pi }\left( \gamma ^{2}\Omega R^{2}c\right) d\varphi
=4\pi \frac{m}{\hbar }\Omega R^{2}\gamma \label{eq:deltaphigab}
\end{equation}
and the proper time difference is

\begin{equation}
\Delta T =4\pi \frac{\Omega R^{2}\gamma }{c^{2}} \equiv \frac{4\pi
R^2 \Omega}{c^2}\left( 1-\frac{\Omega ^{2}R^{2}}{c^{2}}
\right)^{-1/2} \label{eq:deltatau}
\end{equation}

Eq. (\ref{eq:deltatau}) agrees with the well known Sagnac time
delay (see for instance\cite{rizzi03b}).

\subsection{Schwarzschild space-time}\label{ssec:schw}

The standard form of the classical Schwarzschild solution,
describing the vacuum space-time around a spherically symmetric
mass distribution\footnote{See, for instance, \cite{MTW},
Sec.23.6.} is\footnote{In this and in the following Subsections,
we use geometric units such that $G=c=1$. Nevertheless, we keep
$c$ and $G$ in the final results, for the sake of clarity.}
\begin{equation}
ds^2=-\left(1-\frac{2M}{r'} \right)dt'^2+\left(1-\frac{2M}{r'}
\right)^{-1}dr'^2+r'^2\left(d\vartheta'^2+\sin^2 \theta'
d\varphi'^2 \right) \label{eq:sch1}
\end{equation}
where $M$ is the mass of the source, and the coordinate
$(t,r,\vartheta,\varphi)$ are adapted to a congruence of
asymptotically inertial observers. On applying the transformation
(\ref{eq:sagnac_trasf_cord}) to (\ref{eq:sch1}), and setting
$\vartheta=\frac{\pi}{2}$, we get
\begin{equation}
ds^2=-\left(1-\frac{2M}{r}-\Omega^2 r^2 \right)dt^2+ \left(
1-\frac{2M}{r} \right)^{-1}dr^2+r^2d\varphi^2+2\Omega r^2 d\varphi
dt \label{eq:sch2}
\end{equation}
So, if we evaluate along the trajectories $r=R=constant$ the non
null components of the vector field $\bm{\gamma}(x)$ relative to
(\ref{eq:sch2}), we get
\begin{equation}\left\{
\begin{array}{l}
\gamma ^{0} = \gamma_M \\
\gamma _{0} = -\left(\gamma_M\right)^{-1}  \\
\gamma _{\varphi}= \Omega R^2 \gamma_M
\end{array}
\right. \label{eq:gammasatildecurv1}
\end{equation}
and the only non null component of the gravito-magnetic potential
turns out to be
\begin{equation}
\widetilde{A}^G_\varphi  =  -\Omega R^2 \gamma^2_M \label{eq:agm1}
\end{equation}
In both (\ref{eq:gammasatildecurv1}) and (\ref{eq:agm1}) we have
introduced
\begin{equation}
\gamma_M \doteq \left(1-\frac{2M}{R}-\Omega^2 R^2 \right)^{-1/2}
\label{eq:gammam2}
\end{equation}
Hence, the phase shift is given (in physical units) by
\begin{eqnarray}
\Delta \Phi & = & \frac{2m\gamma _{0}}{c\hbar }\oint_{C}{\bf \widetilde{A}}^{G}\cdot {\rm d}%
\widetilde{\bm{r}} \nonumber \\ & = & \frac{2m \gamma_0}{c\hbar
}\int_{0}^{2\pi }\widetilde{A}_{\varphi }^{G}d\varphi \nonumber \\
& = & \frac{4\pi m}{\hbar} \Omega R^2 \gamma_M
\label{eq:sagnacsch1}
\end{eqnarray}
while the corresponding time delay turns out to be
\begin{equation}
\Delta T= \frac{\hbar}{mc^2} \Delta \Phi = \frac{4\pi \Omega
R^2}{c^2} \gamma_M \label{eq:sagnacsch2}
\end{equation}
We see that if $\Omega=0$, i.e. if we measure the propagation time
in a non rotating frame, no Sagnac effect arises. In other words,
the propagation is symmetrical in both directions.

\subsection{Kerr space-time}\label{ssec:sagnac_kerr}

The Kerr solution \cite{kerr63} describes the space-time around a
rotating black-hole or, more generally speaking, around a rotating
singularity. The classical form of this solution is given in
Boyer-Lindquist coordinates \cite{boylin67}
$(t,r,\vartheta,\varphi)$, which are adapted to a congruence of
asymptotically inertial observers:
\begin{eqnarray}
ds^{2} &=&-(1-\frac{2M r'}{\rho ^{2}})dt'^{2}+\frac{\rho ^{2}}{\Delta }%
dr'^{2}+\rho ^{2}d\vartheta'^{2} + \nonumber \\ & & +\left(
r'^{2}+a^{2}+\frac{2M r'a^{2}\sin ^{2}\vartheta' }{\rho
^{2}}\right) \sin ^{2}\vartheta' d\varphi'^{2} -\frac{4M r'a\sin
^{2}\vartheta' }{\rho ^{2}}d\varphi' dt'  \nonumber \\ & &
\label{eq:kerr}
\end{eqnarray}
where  $\rho ^{2}=r'^{2}+a^{2}\cos ^{2}\vartheta' $, and $\Delta
=r'^{2}-2M r'+a^{2}$, $a=J/Mc$,  $J$ is the absolute value of the
angular momentum, and the coordinates are arranged in such a way
that the angular momentum is perpendicular to the equatorial
plane.\footnote{Notice that when $M=a=0$, (\ref{eq:kerr})
expresses the flat space-time metric in Boyer-Lindquist
coordinates.}

By applying the transformation (\ref{eq:sagnac_trasf_cord}), and
then setting $\vartheta=\frac{\pi}{2}$, we obtain
\begin{eqnarray}
ds^2 & = &
-\left(1-\frac{2M}{r}+\frac{4Ma}{r}\Omega-(r^2+a^2+\frac{2Ma^2}{r})\Omega^2
\right)dt^2+\frac{r^2}{\Delta}dr^2+r^2d\vartheta^2 + \nonumber \\
 & & + \left(r^2+a^2+\frac{2Ma}{r} \right)d\varphi^2+\left(-\frac{2Ma}{r}
+(r^2+a^2+\frac{2Ma}{r})\Omega \right)d\varphi dt
\label{eq:kerr_rot}
\end{eqnarray}
Consequently, we obtain the components of the vectors field of the
congruence defining the rotating frame, evaluated along the
trajectories $r=R=constant$:
\begin{eqnarray}
\gamma^0 & = & \gamma_K \nonumber \\
\gamma_0 & = & -(\gamma_K)^{-1} \nonumber \\
\gamma_\varphi & = &
\left(-\frac{2Ma}{r}+(r^2+a^2+\frac{2Ma}{r})\Omega \right)\gamma_K
\label{eq:gamma_kerr1}
\end{eqnarray}
and the corresponding non null component of the gravito-magnetic
potential is
\begin{equation}
\tilde{A}^G_\varphi=-c^2\left(
-\frac{2Ma}{r}+(r^2+a^2+\frac{2Ma}{r})\Omega \right)\gamma^2_K
\label{eq:aphi_kerr1}
\end{equation}
where in both (\ref{eq:gamma_kerr1}) and (\ref{eq:aphi_kerr1}) we
introduced
\begin{equation}
\gamma_K \doteq
\left(1-\frac{2M}{r}+\frac{4Ma}{r}\Omega-(r^2+a^2+\frac{2Ma^2}{r})\Omega^2
\right)^{-1/2} \label{eq:gamma_kappa1}
\end{equation}
Consequently, the phase shift turns out to be (in physical units)
\begin{eqnarray}
\Delta \Phi & = & \frac{2m\gamma _{0}}{c\hbar }\oint_{C}{\bf \widetilde{A}}^{G}\cdot {\rm d}%
\widetilde{\bm{r}} \nonumber \\ & = & \frac{2m \gamma_0}{c\hbar
}\int_{0}^{2\pi }\widetilde{A}_{\varphi }^{G}d\varphi \nonumber \\
& = & \frac{4\pi mc}{\hbar} \left(
-\frac{2GMa}{c^2R}+(R^2+a^2+\frac{2GMa}{c^2R})\frac{\Omega}{c}
\right) \gamma_K \label{eq:Deltaphi_kerr2}
\end{eqnarray}
and the corresponding time delay is
\begin{equation}
\Delta T= \frac{\hbar}{mc^2} \Delta \Phi = \frac{4\pi}{c} \left(
-\frac{2GMa}{c^2R}+(R^2+a^2+\frac{2GMa}{c^2R})\frac{\Omega}{c}
\right) \gamma_K \label{eq:DeltaT_kerr1}
\end{equation}
By explicitly writing $\gamma_K$ we get
\begin{equation}
\Delta T=\frac{4\pi}{c}\frac{\left(
-\frac{2GMa}{c^2R}+(R^2+a^2+\frac{2GMa}{c^2R})\frac{\Omega}{c}
\right)}{\left(1-\frac{2GM}{c^2R}+\frac{4GMa}{c^2R}\frac{\Omega}{c}-(R^2+a^2+\frac{2GMa^2}
{c^2R})\frac{\Omega^2}{c^2} \right)^{1/2}}
\label{eq:DeltaT_kerr2}
\end{equation}

The time delay (\ref{eq:DeltaT_kerr2}) is in agreement with the
results obtained by Tartaglia \cite{tartaglia98}, who studied in
full details the General Relativistic corrections to the Sagnac
effect in Kerr space-time.  Tartaglia also evaluated the
approximations of the time delay (\ref{eq:DeltaT_kerr2}). In the
following Subsection we shall obtain approximated results for the
General Relativistic corrections to the Sagnac effect by starting
from  the weak field solution of Einstein equations around a
rotating mass.

\subsection{The weak field around a rotating mass}\label{ssec:sagnac_curved_wf}

The space-time around a weakly  gravitating object of mass $M$ and
angular momentum $\bm{\vec{J}}$ is given\footnote{See for instance
\cite{MTW},  Sec. 18.1, and also \cite{ruggiero02}.} by
\begin{eqnarray}
ds^2 & = & -\left(1-\frac{2M}{r'}
\right)dt'^2+\left(1+\frac{2M}{r'} \right)\left[dr'^2+r'^2
\left(d\vartheta'^2+\sin^2 \vartheta' d\varphi'^2 \right)\right] +
\nonumber \\ & & -\frac{4J}{r'}\sin^2 \vartheta' d\varphi' dt'
\label{eq:wf1}
\end{eqnarray}
where the spherical coordinates $(t,r,\vartheta,\varphi$) (adapted
to a congruence of a asymptotically inertial observers) have been
arranged in such a way that the angular momentum is orthogonal to
the equatorial plane.

If we apply to the metric (\ref{eq:wf1}) the transformation
(\ref{eq:sagnac_trasf_cord}), after setting
$\vartheta=~\frac{\pi}{2}$, we obtain
\begin{eqnarray}
ds^2 & = & -\left[1-\frac{2M}{r}-\left(1+\frac{2M}{r}
\right)\Omega^2 r^2+\frac{4J\Omega}{r} \right]dt^2+ \nonumber \\
& & + \left(1+\frac{2M}{r} \right)\left[dr^2+r^2
d\varphi^2\right]+ 2\left[\left(1+\frac{2M}{r} \right)\Omega
r^2-\frac{2J}{r} \right] d\varphi dt \nonumber \\ & &
\label{eq:wf2}
\end{eqnarray}
Then, the non null components of the vector field
$\bm{\gamma}(x)$, along the trajectories $r=R=constant$, are
\begin{eqnarray}
\gamma^0 & = & \gamma_J \nonumber \\
\gamma_0 & = & -(\gamma_J)^{-1} \nonumber \\
\gamma_\varphi & = & \left[\left(1+\frac{2M}{R}\right)\Omega
R^2-\frac{2J}{R}  \right]\gamma_J \label{eq:gamma_weak1}
\end{eqnarray}
and the corresponding component of the gravito-magnetic potential
is
\begin{equation}
\widetilde{A}^G_\varphi  =  -\left[\left(1+\frac{2M}{R}
\right)\Omega R^2-\frac{2J}{R} \right] \gamma^2_{J}
\label{eq:agj1}
\end{equation}
In both (\ref{eq:gamma_weak1}) and (\ref{eq:agj1}) we have
introduced
\begin{equation}
\gamma_{J} \doteq \left[1-\frac{2M}{R}-\left(1+\frac{2M}{R}
\right)\Omega^2 R^2+\frac{4J\Omega}{R} \right]^{-1/2}
\label{eq:gammaj2}
\end{equation}
As a consequence, re-introducing physical units, we have
\begin{eqnarray}
\Delta \Phi & = & \frac{2m\gamma _{0}}{c\hbar }\oint_{C}{\bf \widetilde{A}}^{G}\cdot {\rm d}%
\widetilde{\bm{r}} \nonumber \\ & = & \frac{2m\gamma_0}{c\hbar}\int_{0}^{2\pi }\widetilde{A}_{\varphi }^{G}d\varphi \nonumber \\
& = & \frac{4\pi m}{\hbar} \left[\left(1+\frac{2GM}{c^2R}
\right)\Omega R^2-\frac{2GJ}{c^2R} \right] \gamma_{J}
\label{eq:wf4}
\end{eqnarray}
and the corresponding time delay is
\begin{equation}
\Delta T= \frac{\hbar}{mc^2} \Delta \Phi =
\frac{4\pi}{c^2}\left[\left(1+\frac{2GM}{c^2R} \right)\Omega
R^2-\frac{2GJ}{c^2R} \right] \gamma_{J} \label{eq:wf5}
\end{equation}
or, by explicitly writing $\gamma_J$
\begin{equation}
\Delta T =\frac{4\pi}{c^2}\frac{\left[\left(1+\frac{2GM}{c^2R}
\right)\Omega R^2-\frac{2GJ}{c^2R}
\right]}{\left[1-\frac{2GM}{c^2R}-\left(1+\frac{2GM}{c^2R}
\right)\frac{\Omega^2 R^2}{c^2}+\frac{4GJ\Omega}{c^2R}
\right]^{1/2}} \label{eq:wf6}
\end{equation}

In (\ref{eq:wf4}) and (\ref{eq:wf5}) we can distinguish two
contributions: the first one is proportional to the angular
velocity $\Omega$ of the observer, and the other one depends on
the absolute value of the angular momentum of the source $J$.

We see that even when $\Omega=0$ a time difference appears: this
is due to the rotation of the source of the gravitational field
(the same holds in Kerr space-time). In other words
\begin{equation}
\Delta T_J \doteq \frac{4\pi}{c^2}\frac{\left[ -\frac{2GJ}{c^2R}
\right]}{\left[1-\frac{2GM}{c^2R}\right]^{1/2}}
\label{eq:DeltaT_J1}
\end{equation}
is what an asymptotically inertial observer would obtain when
measuring the propagation time for a complete round trip of the
two beams, moving in opposite direction along circular orbits
$r=R=constant$. This time difference corresponds  to the so called
\textit{gravito-magnetic} time delay, which has been obtained by
Stodolski\cite{stodolsky79}, Cohen-Mashhoon\cite{cohen93} in weak
field approximation, and by Tartaglia, by a first order
approximation of the time delay in Kerr space-time.

\section{Conclusions}\label{sec:conclusions}

In the context of natural splitting, the relative formulation of
dynamics can be expressed in terms of gravito-electromagnetic
fields; this analogy, which holds in full theory, leads to the
formulation of the gravito-magnetic Aharonov-Bohm effect. We
showed that the Sagnac effect can be interpreted as a
gravito-magnetic Aharonov-Bohm effect in both flat and curved
space-time, and we exploited this formal analogy for calculating
the General Relativistic corrections to the Sagnac effect, in
stationary and axially symmetric geometries. The results that we
obtained, are in agreement with those available in the literature,
and generalize some old approaches.

The Sagnac effect has a "universal" character: in other words, the
Sagnac time delay is the same, independently of the physical
nature and velocities of  the interfering beams, provided that the
latter are the same, in absolute value, as seen in the rotating
frame. In our formalism this universality is expressed by Eq.
(\ref{eq:deltatau1}), where the mass (or the energy) of the
particles of the beams does not appear, and it is explained as an
effect of the geometrical background in which the beams propagate.
Hence, the geometrical approach clearly points out the universal
character of the effect.

\end{document}